\pdfoutput=1
\documentclass{article}
\usepackage{iclr2027_conference,times}
\usepackage{amsmath,amssymb}
\usepackage{graphicx}
\usepackage{float}
\usepackage{placeins}
\usepackage{etoolbox}
\usepackage{multirow}
\usepackage{wrapfig}
\AtBeginEnvironment{table}{\setlength{\abovecaptionskip}{2pt}\setlength{\belowcaptionskip}{5pt}}
\AtBeginEnvironment{figure}{\setlength{\abovecaptionskip}{3pt}}
\PassOptionsToPackage{hyphens}{url}
\usepackage{hyperref}
\usepackage{url}
\usepackage{xspace}
\newcommand{\methodname}{LatentSift\xspace}
\title{\methodname: Policy-State Filtering for Token-Efficient Verification of Software Engineering Agents}
\author{Yuning Han\textsuperscript{1}, Yangchenchen Jin\textsuperscript{1}, Tyler Jandreau\textsuperscript{2}, Jingwei Sun\textsuperscript{1} \\
\textsuperscript{1}University of Florida \\
\textsuperscript{2}Aeronix}
\iclrfinalcopy
\begin{document}
\maketitle
\lhead{Preprint}
\setlength{\abovedisplayskip}{6pt plus 2pt minus 2pt}\setlength{\belowdisplayskip}{6pt plus 2pt minus 2pt}
\setlength{\abovedisplayshortskip}{2pt plus 2pt}\setlength{\belowdisplayshortskip}{4pt plus 2pt minus 1pt}
\raggedbottom
\begin{abstract}

Test-time scaling improves software engineering agents by generating multiple candidate trajectories and selecting the best one. Verifying and selecting among these long interactions can consume as many tokens as generation itself. Existing hybrid workflows first apply an LLM-based execution-free (EF) verifier to filter candidates before running tests, which adds another model pass over every trajectory. We introduce \methodname, a token-free and execution-free filter that replaces this first stage with hidden states the policy already produces while generating the candidates. It represents each candidate through its reasoning, observation, and function-call states, compares them with positive and negative banks of such states collected from successful and unsuccessful trajectories during policy training, and fuses the resulting distance scores with a learned linear score to retain promising candidates for the execution-based stages. On SWE-bench Verified, across three agents and two policy sizes, \methodname cuts EF-verifier tokens by 66.6--81.0\% and total verification tokens, which include test generation, by 49.1--62.1\% at $K=16$, while hybrid Best@16 matches or improves on each agent's reference workflow, rising from 59.26\% to 60.06\% on DeepSWE-Preview.
\end{abstract}

\section{Introduction}
Efforts to improve LLM reasoning in academia and industry follow two broad directions: training more capable models to solve harder problems, and allocating more computation to each problem at inference time. Training can strengthen the model's reasoning ability before deployment~\citep{deepseekai2026deepseekr1incentivizingreasoningcapability}. The latter approach, known as test-time scaling, seeks better answers by allowing longer reasoning, generating multiple candidates, or searching with verifier feedback~\citep{muennighoff2025s1simpletesttimescaling,wang2023selfconsistencyimproveschainthought,aggarwal2023letssamplestepstep,ICLR2025_1b623663,ICLR2025_8c3caae2}. Rather than relying on a single attempt, these methods spend additional inference compute exploring and selecting solutions of the problem.

\begin{figure}[t]
    \centering
    \includegraphics[width=\linewidth]{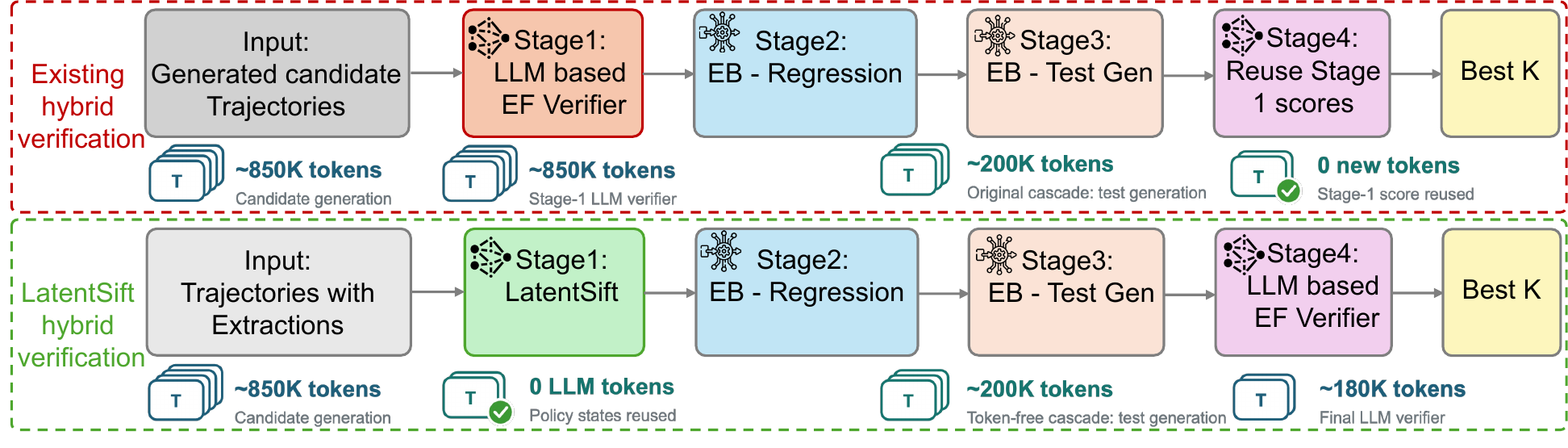}
    \caption{\small\textbf{Existing hybrid verification and the workflow with \methodname.} The existing pipeline follows the hybrid verification design used by R2E-Gym and DeepSWE-Preview~\citep{jain2025r2egymproceduralenvironmentshybrid,deepswe2025}. Both pipelines select a patch from the same pool of $K$ generated candidates. The red and green dashed boxes mark inference-time verification for the existing pipeline and ours, respectively. The existing pipeline scores all candidates with an LLM-based execution-free (EF) verifier in Stage~1 and reuses these scores in Stage~4. \methodname replaces Stage~1 with a filter that reuses cached policy states without additional LLM tokens. Both pipelines retain half the candidates before regression testing and LLM-generated testing; ours applies the final EF verifier only to the remaining candidates. Token counts show average costs per task instance on DeepSWE-Preview at $K=16$, with candidate generation reported separately.}
    \label{fig:hybrid-pipeline}
\end{figure}

For software engineering (SWE) tasks, recent open agents use a hybrid verification workflow to select among generated trajectories~\citep{deepswe2025,jain2025r2egymproceduralenvironmentshybrid}. Figure~\ref{fig:hybrid-pipeline} shows its four stages: an execution-free (EF) verifier scores all candidates and retains half; instance-supplied tests provide regression signals; LLM-generated tests provide additional execution signals; and the EF scores derived from Stage 1 determine the final choice among survivors. The Stage~1 EF verifier is typically a separately fine-tuned LLM that reads the trajectories produced by the policy~\citep{shum2025swermexecutionfreefeedbacksoftware}. Execution-free does not mean token-free: each candidate is processed by the verifier model again, adding verifier input and output tokens. This cost is particularly high on SWE tasks, where a trajectory commonly contains tens of thousands of tokens from reasoning, repository exploration, tool feedback, and code edits. Reprocessing such trajectories adds substantial token consumption to already expensive long-context inference~\citep{kim2026costdynamicreasoningdemystifying,kwon2023efficientmemorymanagementlarge,xie2025wordsaladchopperreasoning}. For a recent open policy such as DeepSWE-Preview with $K=16$ candidates, Stage~1 alone consumes approximately 850K tokens per task instance and the full verification workflow approximately 1,050K, exceeding the approximately 850K tokens used to generate the candidates. Candidate verification can thus rival generation as a source of inference-time token cost.

We introduce a token-efficient verification method called \textbf{\methodname}. It filters candidate trajectories using hidden states collected during the generating process, introducing no additional LLM token cost and requiring no test execution. We represent each trajectory through three channels: reasoning, environment observations, and function calls. For each channel, we compare candidate states with positive and negative banks collected during policy training, aggregate the step-level distance margins, and rank candidates for the task instance. We take the minimum of the three channel ranks as the distance score, so a candidate scores highly only if it ranks highly in every channel. We also train a linear scorer on policy hidden states to assign each candidate a learned trajectory score. We rank these scores within each task instance, normalize the ranks, and average them with the candidates' distance-based scores to obtain the final filtering scores.
This learned signal forms part of the filter. We use \methodname to replace the highly token-costing Stage~1 of the hybrid workflow, retaining half the candidates so execution-based stages and the final LLM verifier operate on a smaller pool, significantly reducing the token cost of the verification workflow.

\textbf{Contributions.} (i) We introduce \methodname, a token-free, execution-free filtering mechanism for long-horizon software engineering agents, and integrate it into an inference-time hybrid verification workflow. (ii) We design a three-channel trajectory representation that separately pools reasoning, observation, and function-call states already computed by the generating policy. We then develop a trajectory scoring method, which combines contrastive distances to successful and unsuccessful experience across the three channels through minimum-rank fusion, then integrates a learned trajectory score to rank candidates within each task instance. (iii) We evaluate \methodname across multiple software engineering agents and policies on SWE-bench Verified, demonstrating substantial verification-token savings while maintaining competitive selection quality. On DeepSWE-Preview, it reduces total verification tokens by 62.1\% while improving hybrid Best@16.

\section{Related Work}
\paragraph{Test-time scaling and agent search.}
Test-time compute can be spent on repeated sampling, adaptive stopping, parallel probing, search, and verifier-guided selection \citep{wang2023selfconsistencyimproveschainthought,aggarwal2023letssamplestepstep,ICLR2024_3fe2a777,zheng2026parallelprobeefficientparallelthinking,ICLR2025_1b623663,ICLR2025_8c3caae2}; on learned allocation and multi-agent pipelines \citep{NEURIPS2025_8d9bbba8,zheng2026llmsimprovingllmsagentic,motwani2025maltimprovingreasoningmultiagent,zhu2025scalingtesttimecomputellm}; or on interaction, revision, and search, as in ReAct, Reflexion, Tree of Thoughts, and language-agent tree search \citep{yao2023reactsynergizingreasoningacting,NEURIPS2023_1b44b878,NEURIPS2023_271db992,zhou2024languageagenttreesearch}. For SWE-bench, CodeMonkeys separates candidate coverage from selection and pays for selection with generated tests and a model-based selector~\citep{ehrlich2025codemonkeys}. We study the marginal cost of that selection when the policy already produces long trajectories.

\paragraph{Software agents and patch verification.}
SWE-bench evaluates generated patches against repository tests \citep{ICLR2024_edac78c3}, and existing systems combine tool-using agents, search, execution signals, and learned execution-free verifiers \citep{NEURIPS2024_5a7c9475,xia2024agentlessdemystifyingllmbasedsoftware,ICLR2025_a1e6783e,jain2025r2egymproceduralenvironmentshybrid,shum2025swermexecutionfreefeedbacksoftware}. SWE-Gym trains agents and verifiers on executable environments and reports Best@$K$ under verifier-guided scaling~\citep{pan2025swegym}, the recipe inherited by the hybrid cascades we build on, whereas SWE-Search spends verifier tokens on LLM value estimates at every search node~\citep{ICLR2025_a1e6783e}. Process supervision, code-agent reward models, and execution-free patch reasoning further strengthen candidate assessment \citep{han2026swetraceoptimizinglonghorizonswe,dihan2026sweshepherdadvancingprmsreinforcing,xu2026scalablesupervisionsoftwareagents}; we target the token cost of applying such assessment at inference time.

\paragraph{Verifiers and process rewards.}
Outcome verifiers and process supervision were developed for mathematical reasoning \citep{cobbe2021trainingverifierssolvemath,ICLR2024_aca97732,uesato2022solvingmathwordproblems,wang2024mathshepherdverifyreinforcellms,ma2023letsrewardstepstep} and extended to Q-value ranking, practical process-reward design, and step-wise promise or progress for agents \citep{ICLR2025_26494b66,choudhury2025processrewardmodelsllm,zhang2025lessonsdevelopingprocessreward,xi2025agentprmprocessrewardmodels}. Preference optimization offers a related route to comparative judgments \citep{NEURIPS2023_a85b405e}, and RewardBench, reward-overoptimization scaling laws, and LLM-as-a-judge studies document the robustness limits of learned evaluators \citep{lambert2024rewardbenchevaluatingrewardmodels,gao2022scalinglawsrewardmodel,NEURIPS2023_91f18a12}. Our verifier pairs a non-parametric distance judgment with a lightweight learned score, both computed from cached policy states.

\paragraph{Signals in internal representations.}
Activations encode truth, confidence, hallucination, and reasoning correctness \citep{azaria2023internalstate,burns2024discoveringlatentknowledgelanguage,kadavath2022languagemodelsmostlyknow,ICLR2025_a712d461,NEURIPS2023_81b83900,marks2024geometrytruthemergentlinear,kossen2024semanticentropyprobesrobust,zhang2025reasoningmodelsknowtheyre,ghasemabadi2026llmspredictfailuresselfawareness}, and distance to expert representations can steer visuomotor policies toward familiar behavior \citep{NEURIPS2025_feaf2359}. We turn such evidence into candidate selection for long-horizon language agents: success and failure banks built during policy training supply outcome-sensitive experience, the three-channel decomposition preserves the structure of agent interaction, and online state collection avoids replaying a trajectory through another language model.

\section{Method}
\label{sec:method}
\subsection{Problem Setup}
\label{sec:setup}

For each task instance $x_m$, a fixed policy generates a pool of $K$ candidate trajectories, $\mathcal{T}_m=\{\tau_{m,i}\}_{i=1}^{K}$. Each trajectory ends in a patch, with $y_{m,i}\in\{0,1\}$ indicating whether it resolves the task under the evaluation protocol. A hybrid selection procedure $\mathcal{S}$ returns a candidate index $i_m^\star=\mathcal{S}(x_m,\mathcal{T}_m)$. As illustrated in Figure~\ref{fig:hybrid-pipeline}, this procedure combines LLM-based execution-free (EF) verification with execution-based (EB) verification using regression and LLM-generated tests. The stages filter candidates using verifier scores and test results, then select a final patch. The selection procedure has no access to the outcome labels or the held-out tests used for final evaluation. Across $M$ task instances, we measure the selected resolution rate $\frac{1}{M}\sum_m y_{m,i_m^\star}$ and compare it with the oracle rate $\frac{1}{M}\sum_m\max_i y_{m,i}$ of the same candidate pools. Our focus is to reduce the additional language-model tokens consumed by hybrid verification while maintaining the selected resolution rate. We denote this cost by $V(\mathcal{S})=N_{\mathrm{in}}+N_{\mathrm{out}}$, counting input and output tokens used by LLM-based EF verification and test generation, and excluding candidate generation.

\subsection{Overview}
Figure~\ref{fig:main} illustrates \methodname, which serves as the
first-stage candidate filter in the hybrid verification workflow
shown in Figure~\ref{fig:hybrid-pipeline}. It replaces the initial LLM-based EF verifier and retains half the candidates for the subsequent regression tests, LLM-generated tests, and final LLM-based EF verification. During policy training, we collect reasoning, observation, and function-call states from trajectories with known outcomes and store them together in positive and negative banks. During inference, we collect candidate states as the policy generates each trajectory. The distance branch compares these states with successful and unsuccessful experience in the banks, aggregates the evidence over steps, and combines the three channel ranks into a distance score. A trained linear scorer provides a learned trajectory score for each candidate. We rank the learned scores within the task instance and average the normalized ranks with the distance scores to obtain the final filtering scores. The filter retains half the candidates without additional LLM tokens or test execution. The subsequent EB stages and final LLM verifier operate on the retained pool. The following sections describe state extraction, bank construction, distance scoring, and fusion with the learned signal.

\begingroup
\setlength{\intextsep}{3pt}
\setlength{\abovecaptionskip}{3pt}
\begin{figure}[!htbp]
    \centering
    \includegraphics[width=1.0\linewidth]{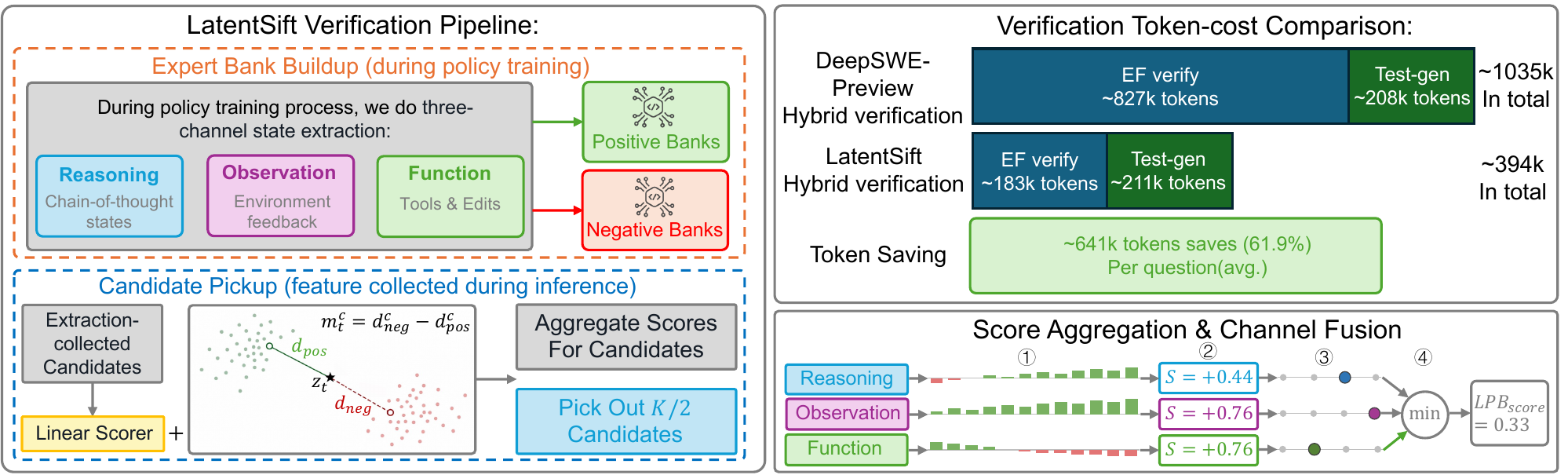}
    \small
    \caption{\small\textbf{\methodname verification design.} \textbf{Left:} Reasoning, observation, and function states from outcome-labeled training trajectories form positive and negative banks. Inference reuses candidate states from ordinary policy forward passes, and Stage~1 averages the distance score with the linear scorer's within-instance rank to retain the top half of candidates. \textbf{Lower right:} Distance scoring: (1) retrieve the nearest positive and negative states and compute $m_t^c=d_{\mathrm{neg}}^c-d_{\mathrm{pos}}^c$. (2) aggregate by step-weighted mean; (3) rank within each task instance and channel; (4) take the minimum channel rank. \textbf{Upper right:} Average verification tokens per task instance for the original and token-free cascades at $K=16$ (Section~\ref{sec:cross-agent}).}
    \label{fig:main}
\end{figure}
\endgroup

\subsection{Three-channel representations and bank construction}

We represent each trajectory through three types of spans:
\textbf{reasoning}, the policy's analysis outside function calls;
\textbf{observation}, environment feedback such as retrieved code,
command output, and test results; and \textbf{function}, tool names
and arguments, including commands and code edits.
We identify these spans using message roles and function-call
boundaries. For trajectory $\tau_i$, let
$h_{i,p}^{(L)}\in\mathbb{R}^{d}$ denote the final-layer state at
token position $p$. At step $t$, we mean-pool the states assigned
to channel $c$:
\begin{equation}
z_{i,t}^{c}
=
\frac{1}{|\mathcal{S}_{i,t}^{c}|}
\sum_{p\in\mathcal{S}_{i,t}^{c}} h_{i,p}^{(L)},
\qquad
c\in\{\mathrm{cot},\mathrm{obs},\mathrm{fn}\},
\end{equation}
where $\mathcal{S}_{i,t}^{c}$ contains the token positions of that
channel. The task description and structural delimiters are
excluded from pooling.

\begin{wrapfigure}{r}{0.42\linewidth}
    \vspace{-12pt}
    \centering
    \includegraphics[width=\linewidth]{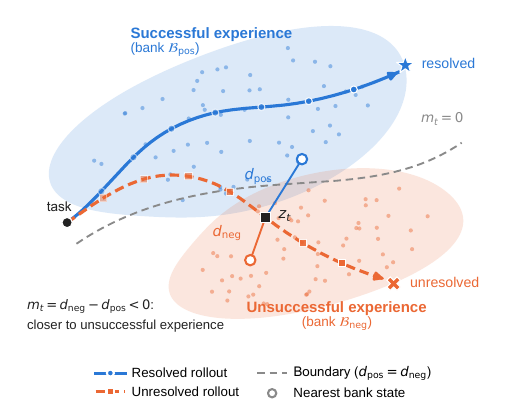}
    \vspace{-16pt}
    \caption{\small\textbf{Contrastive distance} of a candidate state to the successful and unsuccessful banks, with margin $m_t=d_{\mathrm{neg}}-d_{\mathrm{pos}}$.}
    \label{fig:lpb-schematic}
    \vspace{-10pt}
\end{wrapfigure}
During policy training, we collect these representations from
trajectories with executable outcome labels. Trajectories that
resolve their tasks form the positive bank
$\mathcal{B}_{\mathrm{pos}}$, and unsuccessful trajectories form
the negative bank $\mathcal{B}_{\mathrm{neg}}$.
Each bank entry stores the reasoning, observation, and function
state sequences of one trajectory together.
The banks are built from training repositories that do not overlap with
the evaluation repositories.

During inference, we collect the same three types of
representations from candidate trajectories as they are generated.
Training representations come from the policy being trained,
whereas candidate representations come from the policy used for
inference, and both use final-layer states.
These candidate states are then compared with the corresponding
channel states in the positive and negative banks, as described
in the next section.

\subsection{Distance scoring and latent filtering}
\label{sec:distance-gate}
For a candidate state $z_{i,t}^{c}$ we retrieve its nearest positive and negative bank states in the same channel (Figure~\ref{fig:lpb-schematic}), writing $\mathcal{Z}^{c}(\mathcal{B})$ for the channel-$c$ states of bank $\mathcal{B}$: $d_{\mathrm{pos}}^{c}(z)=\min_{u\in\mathcal{Z}^{c}(\mathcal{B}_{\mathrm{pos}})}\|z-u\|_2$ and $d_{\mathrm{neg}}^{c}(z)=\min_{u\in\mathcal{Z}^{c}(\mathcal{B}_{\mathrm{neg}})}\|z-u\|_2$. The signed margin is aggregated over steps with position weights,
\begin{equation}
 m_{i,t}^{c}=d_{\mathrm{neg}}^{c}(z_{i,t}^{c})-d_{\mathrm{pos}}^{c}(z_{i,t}^{c}),
 \qquad q_i^{c}=\frac{\sum_{t=0}^{T_i-1}t\,m_{i,t}^{c}}{\sum_{t=0}^{T_i-1}t},
 \label{eq:distance-margin}
\end{equation}
so that later steps, taken after environment feedback, count more. The positive value indicates proximity to successful rather than unsuccessful experience. Within each task instance and channel, we rank candidates by $q_i^{c}$ in ascending order, average the ranks of tied scores, and scale the zero-based ranks to $[0,1]$. Higher scores receive higher ranks $r_i^{c}$ (Appendix~\ref{app:rank-normalization}). We take $s_{\mathrm{dist}}(\tau_i)=\min_{c}r_i^{c}$, so a candidate scores highly only if every channel ranks it highly. Appendix~\ref{app:distance-gate-details} gives the rationale and cost accounting, and Section~\ref{sec:ablation} ablates channel subsets, one-sided distances, and naive filtering rules including uniform step weights and trajectory length.

\subsection{Linear scorer and signal fusion}
\label{sec:linear-scorer}
A lightweight linear scorer inspired by SWIFT~\citep{guo2026swift} complements the distance judgment. Trained on the same outcome-labeled trajectories, it reads pooled reasoning and function states from all transformer layers, excluding observations. A shared linear head maps each span to a gate logit and a local reward, and their gated average is the trajectory score, trained with binary cross-entropy against the outcome (Appendix~\ref{app:learned-scorer}). In inference, it reads only cached candidate states. We rank the learned trajectory scores within each task instance using the normalization in Appendix~\ref{app:rank-normalization}, and denote the resulting score for candidate $\tau_i$ by $s_{\mathrm{lin}}(\tau_i)\in[0,1]$. We combine the two branch scores with equal weights,
\begin{equation}
 s(\tau_i)=0.5\,s_{\mathrm{dist}}(\tau_i)+0.5\,s_{\mathrm{lin}}(\tau_i),
 \label{eq:scorer-fusion}
\end{equation}
and the $\max(3,\lfloor K/2\rfloor)$ highest-scoring candidates, the top half at $K=16$, proceed to the execution-based stages and the final EF verifier.

\section{Experiments}
\label{sec:experiments}
\subsection{Experimental setup}
\label{sec:experimental-setup}

\paragraph{Task.}

We evaluate on the 500 human-validated GitHub issues in SWE-bench Verified~\citep{ICLR2024_edac78c3}. For each task instance, an agent generates multiple trajectories, each ending in a candidate repository patch. The verification workflow evaluates these candidates and selects one patch. We determine whether the selected patch resolves the instance using the official evaluation. All verification methods use the same candidate pool within each agent-policy configuration, so differences in performance reflect candidate verification rather than generation.

\paragraph{Agents, policies, and candidate pools.}
We evaluate \methodname on multiple agents, including DeepSWE Agent, R2EGym Agent~\citep{jain2025r2egymproceduralenvironmentshybrid}, and CWM Agent~\citep{copet2025cwm}. DeepSWE Agent denotes the official DeepSWE-Preview~\citep{deepswe2025} agentic scaffold, comprising its prompt template, tool protocol, and agent configuration, and R2EGym Agent denotes the official R2E-Gym scaffold.\footnote{\href{https://github.com/R2E-Gym/R2E-Gym/blob/main/reproduction/DEEPSWE_REPRODUCTION.MD}{Official DeepSWE reproduction guide.}} Both are implemented in the R2E-Gym framework and share its editing interface, but they are distinct agent configurations. The verification workflows of DeepSWE and R2EGym are similar, both composed of the four stages shown in Figure~\ref{fig:hybrid-pipeline}. For CWM, since its test-time scaling verifier is not open-sourced, we reuse the verification workflow of DeepSWE Agent.

We also conduct evaluation on various policy models. DeepSWE-Preview is RL-trained from Qwen3-32B~\citep{deepswe2025}, R2EGym-32B is fine-tuned from Qwen2.5-Coder-32B-Instruct~\citep{jain2025r2egymproceduralenvironmentshybrid}, and CWM is a 32B open-weights model~\citep{copet2025cwm}. We compare the performance on the same agent across different policy models by evaluating on R2EGym Agent with R2EGym-32B and R2EGym-14B~\citep{jain2025r2egymproceduralenvironmentshybrid} models. Detailed checkpoint identities and implementation choices are provided in Appendix~\ref{app:verifier-configs}.

For each policy--agent configuration, we generate 16 trajectories for each of the 500 task instances, yielding 8,000 trajectories per configuration. Verifier comparisons use the same pool. Changing the agent scaffold requires generating new trajectories and running the tests on their patches again.
Following prior hybrid workflows~\citep{deepswe2025,jain2025r2egymproceduralenvironmentshybrid},
the R2E-TestgenAgent~\citep{r2egym2025testgen} was employed to generate one set of tests per task instance.
For each agent, we build positive and negative banks from trajectories and their three-channel hidden states collected during generator model training, containing 1,383 successful and 1,383 unsuccessful trajectories from repositories disjoint from the evaluation repositories. For each $K$, we report mean Best@$K$ over 200 draws of $K$ candidates from each fixed pool, using stored test results.

\paragraph{Verifier baselines.}
We compare our \methodname with multiple EF verifiers, including DeepSWE-Verifier~\citep{agentica2025deepsweverifier}, R2EGym-Verifier~\citep{r2egym2025verifier}, DEV matching-pairs~\citep{r2edits2025devmatchingpairs} and AgentPRM~\citep{xi2025agentprmprocessrewardmodels}. Each baseline uses its own verifier as the EF verifier in the agent's verification workflow (i.e., Stages~1 and~4 in Figure~\ref{fig:hybrid-pipeline}). Ours uses \methodname in Stage~1 and the policy-associated EF verifier in Stage~4. Each method retains $\max(3,\lfloor K/2\rfloor)$ candidates, the top half at $K=16$.

\paragraph{Metrics.}
Our primary quality metric is Best@$K$, the percentage of task instances solved by the candidate selected from a pool of size $K$; hybrid Best@$K$ is this rate after the complete four-stage cascade. Oracle Pass@$K$ reports the corresponding candidate-generation ceiling by assuming that the best candidate will always be selected. We measure verification token cost by reporting EF-verifier tokens and total verification tokens including test generation. EF-verifier token counts use the calls made by each workflow, reusing Stage~1 scores in Stage~4 when available. We charge the logged test-generation cost to a workflow only if it reaches Stage~3.

\subsection{Main results}
\label{sec:main-results}
We evaluate hybrid verification across agent configurations and policy sizes. The cross-agent experiment compares four baseline EF verifiers with \methodname, each baseline serving in both EF stages and ours paired with the policy-associated final verifier. The cross-policy experiment compares each policy's own verifier with \methodname under a fixed scaffold.

\subsubsection{Cross-agent evaluation}
\label{sec:cross-agent}

We compare DeepSWE Agent (DeepSWE-Preview), R2EGym Agent (R2EGym-32B), and CWM Agent (CWM). CWM ships no verifier, so we retrained four verifiers on Llama-3.1-8B-Instruct with the baselines' original recipes.

\begin{table}[!htbp]
\caption{Cross-agent hybrid results on SWE-bench Verified at $K=16$. Best@16 is the resolution rate (\%). Token cost gives EF-verifier (EF) and total verification tokens per task instance, in thousands.}
\label{tab:cross-agent-main}
\centering
\footnotesize
\setlength{\tabcolsep}{2.5pt}
\begin{tabular}{lrrrrrrrrr}
\hline\noalign{\vskip 1.5pt}
& \multicolumn{3}{c}{DeepSWE Agent} & \multicolumn{3}{c}{R2EGym Agent} & \multicolumn{3}{c}{CWM Agent}\\
\multirow{2}{*}{Stage~1 verifier} & \multirow{2}{*}{Best@16} & \multicolumn{2}{c}{Token cost (K)} & \multirow{2}{*}{Best@16} & \multicolumn{2}{c}{Token cost (K)} & \multirow{2}{*}{Best@16} & \multicolumn{2}{c}{Token cost (K)}\\
\cline{3-4}\cline{6-7}\cline{9-10}\noalign{\vskip 0.8pt}
& & EF & Total & & EF & Total & & EF & Total\\
\hline\noalign{\vskip 1.5pt}
DeepSWE-Verifier & 59.26 & 826.7 & 1035.1 & 49.70 & 367.4 & 599.4 & 54.21 & 692.6 & 930.0\\
R2EGym-Verifier & 57.95 & 774.1 & 982.0 & 46.56 & 366.2 & 599.6 & 53.51 & 707.1 & 944.6\\
DEV matching-pairs & 58.75 & 826.7 & 1037.6 & 48.38 & 367.4 & 599.4 & 53.11 & 692.6 & 925.6\\
AgentPRM & 59.76 & 810.9 & 1022.8 & 46.56 & 358.3 & 586.8 & 53.31 & 677.3 & 909.8\\
\textbf{Ours} & \textbf{60.06} & \textbf{181.0} & \textbf{391.9} & \textbf{47.73} & \textbf{76.0} & \textbf{304.9} & \textbf{54.21} & \textbf{231.3} & \textbf{466.8}\\
\noalign{\vskip 1pt}\hline
\end{tabular}
\end{table}

Table~\ref{tab:cross-agent-main} reports hybrid Best@16 and verification cost for the five Stage-1 verifiers under the protocol of Section~\ref{sec:experimental-setup}. For DeepSWE, no baseline verifier exceeds \methodname (60.06\%), and on CWM Agent \methodname matches the strongest baseline (54.21\%) with a third of its EF tokens (231.3K against 692.6K). For the R2EGym-32B pool, \methodname raises Best@16 by 1.17 points over the original R2EGym-Verifier cascade while cutting EF tokens by 4.8$\times$ and total verification cost by half. Additional Best@$K$ results for $K=4,8,12$ appear in Appendix~\ref{app:additional-results}.

\begingroup
\setlength{\intextsep}{3pt}
\setlength{\abovecaptionskip}{3pt}
\begin{figure}[!htbp]
    \centering
    \includegraphics[width=0.95\linewidth]{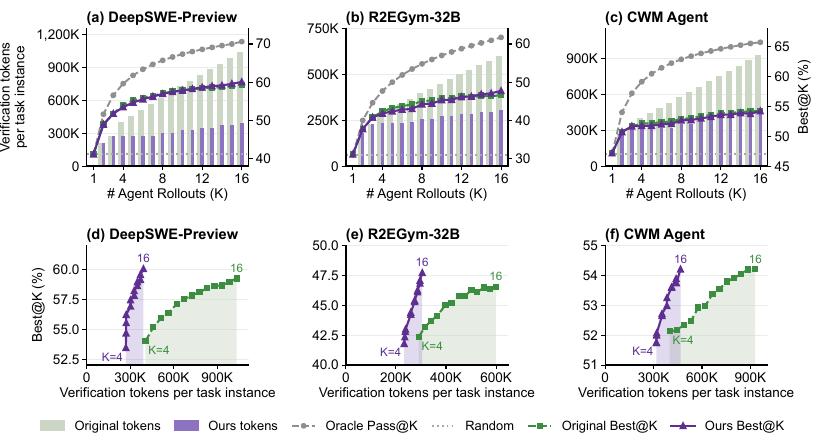}
    \small
    \caption{\small\textbf{Verification cost and resolution rate on the three agents.} Top: verification tokens per task instance, including test generation (bars, left axis), and Best@$K$ with Oracle Pass@$K$ and random selection (curves, right axis) as $K$ grows. Bottom: the same runs as Best@$K$ against verification tokens for $K$ from 4 to 16. The baseline hybrid uses the policy-associated verifier in Stages~1 and~4, and ours replaces it in Stage~1 with \methodname.}
    \label{fig:main-results}
\end{figure}
\endgroup

Figure~\ref{fig:main-results} (top) shows that \methodname tracks the Best@$K$ of the existing hybrid at every candidate budget on all three agents while its verification cost grows more slowly, so the token saving widens with $K$. At $K=16$ it improves Best@16 over the existing hybrid on DeepSWE-Preview and R2EGym-32B and matches it on CWM Agent, while reducing total verification tokens by 62.1\%, 49.1\%, and 49.8\%, respectively. The bottom row plots the same runs as Best@$K$ against verification tokens: at a matched verification cost, \methodname reaches a higher Best@$K$ than the original cascade on all three agents.

\subsubsection{Cross-policy evaluation}
\label{sec:cross-policy}
\begin{figure}[!htbp]
    \centering
    \includegraphics[width=0.81\linewidth, trim=0 0 221 0, clip]{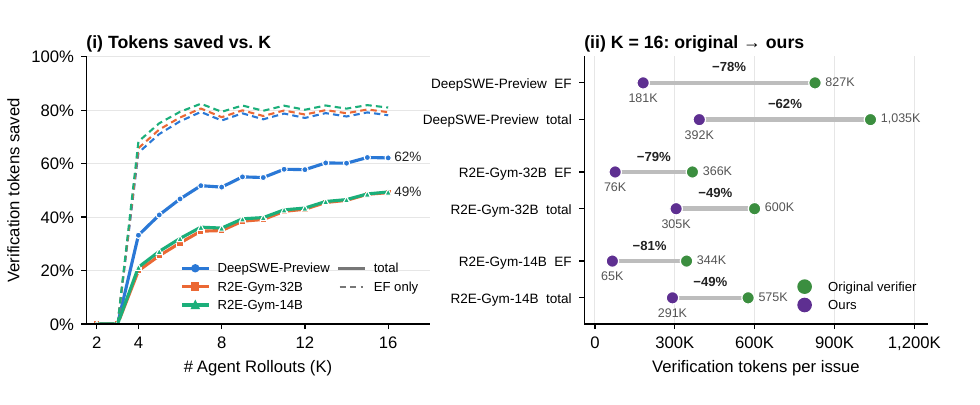}
    \caption{\small\textbf{Verification tokens saved across policy--agent configurations.} The cross-policy comparison holds R2EGym Agent fixed and varies the policy between R2EGym-32B and R2EGym-14B. (i)~Rate of verification tokens saved as $K$ grows (solid: total including test generation; dashed: EF-verifier tokens only). (ii)~Tokens per task instance at $K=16$ with the policy-associated verifier and with \methodname in Stage~1.}
    \label{fig:cross-policy-saving}
\end{figure}

We hold R2EGym Agent fixed and compare R2EGym-32B with R2EGym-14B to conduct evaluation across policy sizes. Each policy supplies its own candidate pool, bank, and linear scorer. Both use R2EGym-Verifier in the original cascade's Stages~1 and~4 and in our Stage~4. Figure~\ref{fig:cross-policy-saving} also includes DeepSWE-Preview with DeepSWE Agent as a reference from the cross-agent comparison in Section~\ref{sec:cross-agent}.

Figure~\ref{fig:cross-policy-saving} shows that \methodname substantially reduces verification tokens on both R2EGym policies while maintaining hybrid resolution rates. At $K=16$, EF-verifier tokens fall from 366K to 76K per task instance on R2EGym-32B and from 344K to 65K on R2EGym-14B. Including test generation, total verification cost falls by about 49\% on each policy, from 600K to 305K and from 575K to 291K, respectively (Appendix~\ref{app:r2e-token-accounting}). Hybrid Best@16 rises from 46.56\% to 47.73\% on 32B and remains within 0.33 percentage points of the original cascade on 14B (Table~\ref{tab:cross-policy-best16}, Appendix~\ref{app:additional-results}). The EF-token saving reaches about 80\% by $K\approx6$ on both policies and persists as the candidate budget grows.

\subsubsection{Score distributions}
\label{sec:score-distributions}

To show where the discrimination of \methodname comes from, we randomly sample 100 of the 500 task instances and plot the scores of their DeepSWE-Preview candidates in Figure~\ref{fig:margin-hist}. In every channel, resolved candidates concentrate at positive distance margins (Equation~\ref{eq:distance-margin}), the successful side of the decision boundary in Figure~\ref{fig:lpb-schematic}, whereas unresolved candidates shift to negative margins with a long tail. The final score $s$ (Equation~\ref{eq:scorer-fusion}, panel d) separates the two groups further, so keeping the top half of each pool discards mainly unresolved candidates.

\begingroup
\setlength{\intextsep}{3pt}
\setlength{\abovecaptionskip}{3pt}
\begin{figure}[!htbp]
    \centering
    \includegraphics[width=0.9\linewidth]{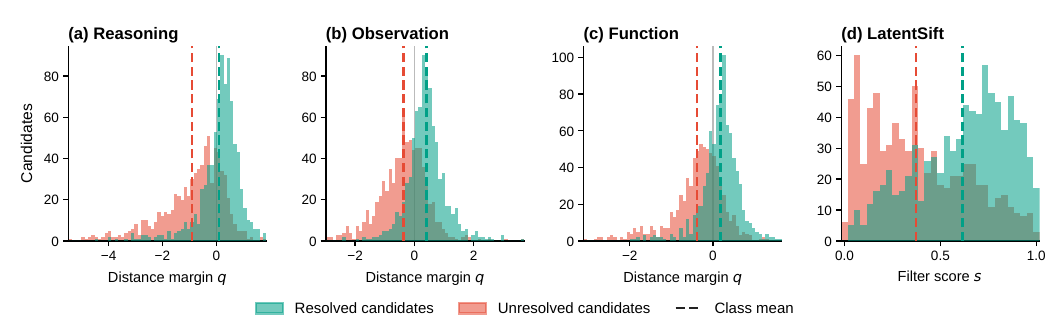}
    \caption{\small\textbf{Score distributions of resolved and unresolved DeepSWE-Preview candidates} on 100 randomly sampled task instances. (a)--(c)~Step-weighted distance margin $q^{c}$ in the reasoning, observation, and function channels. (d)~Final \methodname score $s$. Dashed lines mark class means.}
    \label{fig:margin-hist}
\end{figure}
\endgroup

\subsection{Ablation study}
\label{sec:ablation}

We first isolate the Stage~1 filter through its oracle retention (Section~\ref{sec:gate-retention}), then ablate three design choices of the distance branch at $K=16$, with the policy, scaffold, banks, execution signals, linear scorer, and Stage~4 verifier fixed within each configuration: channel selection, distance construction, and naive filtering rules. Channel subsets take the minimum rank over the selected channels and are fused with the linear scorer as in the full method.

\subsubsection{Stage-1 retention}
\label{sec:gate-retention}

Best@16 credits the Stage~1 filter together with three downstream stages. To isolate the filter we measure its \emph{oracle retention}: among task instances whose $K$ candidates contain a correct patch, the fraction whose retained $\max(3,\lfloor K/2\rfloor)$ candidates still contain one.
Table~\ref{tab:gate-retention} reports it on DeepSWE-Preview at even $K$ for the LLM EF verifier, \methodname, and random filtering.
At $K=16$ \methodname retains a correct candidate on 97.72\% of solvable task instances, against 97.44\% for the LLM verifier and 93.19\% for random filtering. For $K\le8$ the LLM verifier leads by less than one point. The difference narrows as $K$ grows, falls to 0.07 point at $K=12$, and favors \methodname from $K=14$ on. These results show that \methodname retains correct candidates at a rate comparable to the LLM EF verifier, without additional LLM tokens.

\begin{table}[!htbp]
\caption{Oracle retention (\%) after Stage~1 on DeepSWE-Preview at even $K$, retaining $\max(3,K/2)$ candidates, over the same 200 draws as Figure~\ref{fig:main-results}.}
\label{tab:gate-retention}
\centering
\footnotesize
\setlength{\tabcolsep}{6pt}
\begin{tabular}{lrrrrrrr}
\hline\noalign{\vskip 1.5pt}
Stage-1 filter & $K{=}4$ & $K{=}6$ & $K{=}8$ & $K{=}10$ & $K{=}12$ & $K{=}14$ & $K{=}16$\\
\hline\noalign{\vskip 1.5pt}
LLM EF verifier & 97.67 & 94.67 & 95.79 & 96.37 & 96.81 & 97.13 & 97.44\\
\textbf{\methodname (ours)} & 97.07 & 93.77 & 94.95 & 95.97 & 96.74 & 97.28 & 97.72\\
Random filtering & 92.31 & 85.91 & 88.27 & 90.03 & 91.49 & 92.50 & 93.19\\
\noalign{\vskip 1pt}\hline
\end{tabular}
\end{table}

\subsubsection{Channel selection}

\begin{table}[!htbp]
\caption{Channel selection ablation of \methodname at $K=16$: hybrid Best@16 (\%) with the distance branch restricted to subsets of the reasoning (R), observation (O), and function (F) channels.}
\label{tab:channel-ablation}
\centering
\footnotesize
\setlength{\tabcolsep}{5pt}
\begin{tabular}{lccccccc}
\hline\noalign{\vskip 1.5pt}
Distance channels & \textbf{R+O+F} & R+O & R+F & O+F & R & O & F\\
\hline\noalign{\vskip 1.5pt}
DeepSWE Agent & \textbf{60.06} & 59.26 & 59.66 & 59.86 & 59.46 & 59.26 & 59.46\\
R2EGym Agent & \textbf{47.73} & 46.60 & 46.90 & 47.10 & 46.71 & 46.46 & 47.25\\
\noalign{\vskip 1pt}\hline
\end{tabular}
\end{table}

Restricting the distance branch to single channels or pairs, with the learned signal and downstream stages fixed, lowers Best@16 in every case (Table~\ref{tab:channel-ablation}). On DeepSWE Agent the best pair, observation + function, comes closest (59.86\% against 60.06\%) and single channels lose more; on R2EGym Agent the function channel alone is the strongest subset (47.25\% against 47.73\%), ahead of every pair, so the relative value of the channels differs by agent. The three channels therefore carry complementary information.

\subsubsection{Distance construction}
\label{sec:distance-construction}

We evaluate the benefit of using distances to both successful and unsuccessful experience. Our two-sided distance score,
$m_{i,t}^{c}=d_{\mathrm{neg}}^{c}-d_{\mathrm{pos}}^{c}$ (Equation~\ref{eq:distance-margin}), combines proximity to successful trajectories with separation from unsuccessful trajectories. We compare it with positive-only scoring using $d_{\mathrm{pos}}^{c}$ (smaller is better) and negative-only scoring using $d_{\mathrm{neg}}^{c}$ (larger is better), keeping all other components fixed.

\begin{table}[!htbp]
\caption{Distance construction ablation at $K=16$ (\%): hybrid Best@16 and oracle retention (Ret.).}
\label{tab:distance-ablation}
\centering
\footnotesize
\setlength{\tabcolsep}{8pt}
\begin{tabular}{lrrrr}
\hline\noalign{\vskip 1.5pt}
& \multicolumn{2}{c}{DeepSWE Agent} & \multicolumn{2}{c}{R2EGym Agent}\\
\cline{2-3}\cline{4-5}\noalign{\vskip 0.8pt}
Distance score & Best@16 & Ret. & Best@16 & Ret.\\
\hline\noalign{\vskip 1.5pt}
\textbf{$d_{\mathrm{neg}}^c-d_{\mathrm{pos}}^c$} & \textbf{60.06} & \textbf{97.72} & \textbf{47.73} & \textbf{92.13}\\
$d_{\mathrm{pos}}^c$ & 58.05 & 97.30 & 44.16 & 89.31\\
$d_{\mathrm{neg}}^c$ & 59.86 & 97.44 & 47.30 & 91.48\\
\noalign{\vskip 1pt}\hline
\end{tabular}
\end{table}

Two-sided distance scoring achieves the highest hybrid Best@16 and oracle retention on both agents (Table~\ref{tab:distance-ablation}). Compared with the one-sided alternatives, it improves Best@16 by 0.20--2.01 percentage points on DeepSWE Agent and 0.43--3.57 points on R2EGym Agent. These results support using both successful and unsuccessful experience for candidate filtering. We also observe an interesting pattern: negative-only scoring
outperforms positive-only scoring, suggesting that unsuccessful experience may provide more informative evidence for candidate filtering.

\subsubsection{Naive filtering rules}
\label{sec:naive-filters}

Finally, we test whether step weighting is necessary and how naive rules that bypass the three-channel representation perform. The uniform step mean $q_{i,\mathrm{uniform}}^{c}=\frac{1}{T_i}\sum_{t}m_{i,t}^{c}$ drops only the step weighting and stays fused with the linear scorer. Two length rules, preferring shorter trajectories in tokens or agent steps, serve alone as the Stage-1 filter. Stages~2--4 are unchanged.

\begin{table}[!htbp]
\caption{Naive Stage-1 filtering rules against \methodname at $K=16$ (\%): hybrid Best@16 and oracle retention (Ret.).}
\label{tab:naive-filters}
\centering
\footnotesize
\setlength{\tabcolsep}{8pt}
\begin{tabular}{lrrrr}
\hline\noalign{\vskip 1.5pt}
& \multicolumn{2}{c}{DeepSWE Agent} & \multicolumn{2}{c}{R2EGym Agent}\\
\cline{2-3}\cline{4-5}\noalign{\vskip 0.8pt}
Stage-1 rule & Best@16 & Ret. & Best@16 & Ret.\\
\hline\noalign{\vskip 1.5pt}
\textbf{\methodname (step-weighted mean)} & \textbf{60.06} & \textbf{97.72} & \textbf{47.73} & \textbf{92.13}\\
Uniform step mean & 59.66 & 97.44 & 46.36 & 90.16\\
Trajectory length, tokens & 57.24 & 94.87 & 45.65 & 91.15\\
Trajectory length, steps & 58.39 & 95.29 & 47.60 & 92.40\\
\noalign{\vskip 1pt}\hline
\end{tabular}
\end{table}

\methodname achieves the highest Best@16 on both agents (Table~\ref{tab:naive-filters}), outperforming the three naive rules by 0.40--2.82 points on DeepSWE Agent and by 0.13--2.08 points on R2EGym Agent. The relative performance of these naive rules varies across agents, whereas \methodname consistently achieves the highest performance on both agents.

\section{Conclusion and Future Work}
We presented \methodname, a token-free and execution-free filter that scores candidate trajectories from the policy's own reasoning, observation, and function-call states instead of replaying them through a second language model. Across three agents at $K=16$, \methodname cuts EF-verifier tokens by 66.6--79.2\% and total verification cost by 49.1--62.1\% while raising or matching hybrid Best@16, and the savings carry over to a smaller policy and grow with the candidate budget. Policy-state filtering is therefore a drop-in replacement for the most token-consuming stage of hybrid verification.

\methodname applies to any policy whose hidden states are accessible, including the open-weight agents studied here. Our evaluation covers candidate budgets up to $K=16$ on the Python repositories of SWE-bench Verified. Extending it to larger budgets, other programming languages and benchmarks, and larger experience banks is a natural next step and may improve performance-efficiency tradeoffs further.

\clearpage
\bibliographystyle{iclr2027_conference}
\bibliography{references}

\clearpage
\appendix
\section{Implementation and Reporting Details}
\label{sec:appendix}

\subsection{Distance aggregation, selection, and cost accounting}
\label{app:distance-gate-details}
\paragraph{Banks and retrieval.}
Each bank entry stores the three channel sequences of one trajectory, so channel-$c$ retrieval searches the channel-$c$ states of all bank trajectories and no separate per-channel bank is built. The linear scorer is trained on the same trajectories (Appendix~\ref{app:learned-scorer}). Neither signal uses evaluation outcomes.

\paragraph{Contrastive distance.}
The step margin $m_{i,t}^{c}$ of Equation~\ref{eq:distance-margin} is positive when the nearest successful bank state is closer to the candidate state than the nearest unsuccessful one. A one-sided distance measures proximity to a single outcome, while the margin weighs the two outcomes against each other. Section~\ref{sec:distance-construction} compares the three constructions.

\paragraph{Temporal and channel aggregation.}
The weight of step $t$ grows linearly with $t$, so evidence from later steps, taken after the agent has received environment feedback and revised its solution, counts more. Section~\ref{sec:naive-filters} evaluates the uniform step mean. The three channel scores differ in scale and distribution, so summing them directly would let the channel with the largest magnitude dominate. Within-instance ranks place the channels on a common scale without requiring calibrated success probabilities. Taking the minimum rank sets the distance score to the rank of the weakest channel, so ranks of $0.67$, $1.00$, and $0.33$ give $s_{\mathrm{dist}}=0.33$. No separate rejection threshold is needed, and the learned score enters afterwards through Equation~\ref{eq:scorer-fusion}.

\paragraph{Within-instance rank normalization.}
\label{app:rank-normalization}
For each task instance and channel $c$, let $n$ be the number of candidates being ranked. We sort their scores $q_i^c$ in ascending order and assign zero-based ranks $0,\ldots,n-1$, giving tied candidates the mean of the ranks they occupy. Writing this rank as $\rho_i^c$, we compute
\begin{equation}
 r_i^c=\frac{\rho_i^c}{\max(n-1,1)},
 \qquad s_{\mathrm{dist}}(\tau_i)=\min_c r_i^c.
\end{equation}
Ranks are never pooled across task instances. Without ties and for $n>1$, the lowest and highest scores map to 0 and 1. If all $n>1$ scores are equal, every candidate receives rank $0.5$, and a single candidate receives rank 0. For example, scores $(1,1,2,2)$ yield normalized ranks $(1/6,1/6,5/6,5/6)$. Averaging tied ranks makes the result independent of the order of equal-scoring candidates.

\paragraph{Candidate selection.}
In the hybrid cascade, \methodname retains the $\max(3,\lfloor K/2\rfloor)$ candidates with the highest fused scores $s(\tau_i)$. The execution-based stages prune this set further, and the final EF verifier scores only the candidates that survive them (Appendix~\ref{app:r2e-token-accounting}).

\paragraph{Cost accounting.}
Token-free means that the filter adds no LLM input or output tokens, not that it requires no computation. At inference, the filter mean-pools states that the policy's forward passes already produce, retrieves nearest neighbors from the two banks in each channel, and applies one linear head. Bank construction and scorer training are performed offline. A text verifier that rereads $K$ trajectories processes roughly $K$ times the average trajectory length in input tokens before prompt overhead, so the verification tokens of the original cascade grow linearly with $K$ (Figure~\ref{fig:main-results}). We count EF-verifier calls and test-generation calls separately, include only calls that are actually made, and exclude candidate generation. In the original cascade, Stage~4 reuses the Stage-1 EF scores and makes no new verifier calls (Appendix~\ref{app:verifier-configs}).

\subsection{EF verifier configurations}
\label{app:verifier-configs}
DeepSWE-Verifier and R2EGym-Verifier are the official verifiers of DeepSWE-Preview and R2E-Gym. They are LoRA adapters on Qwen3-14B and Qwen2.5-Coder-14B-Instruct, respectively. DEV matching-pairs is another Qwen3-14B LoRA adapter, trained on matched pairs of one successful and one unsuccessful rollout of the same task. All three adapters use rank 64 and $\alpha=128$~\citep{agentica2025deepsweverifier,r2egym2025verifier,r2edits2025devmatchingpairs}. The DeepSWE-Verifier and DEV matching-pairs model cards report a learning rate of $10^{-5}$, two training epochs, and cosine learning-rate decay with a warmup ratio of 0.05. AgentPRM~\citep{xi2025agentprmprocessrewardmodels} was proposed for web agents such as web shopping and browser navigation. We port its official implementation to coding-agent trajectories and train it as a LoRA adapter with the same base model and configuration as DeepSWE-Verifier. CWM releases no verifier, so for CWM Agent we retrain the four baseline verifiers on Llama-3.1-8B-Instruct with their original recipes.

In the official hybrid workflow, the final stage ranks the candidates that pass the regression and generated tests by their stored Stage-1 EF scores and calls no separate verifier~\citep{deepswe2025,jain2025r2egymproceduralenvironmentshybrid}. Each EF baseline therefore uses its own checkpoint in Stages~1 and~4. \methodname replaces Stage~1 and calls the policy-associated verifier only in Stage~4, namely DeepSWE-Verifier for DeepSWE-Preview, R2EGym-Verifier for both R2EGym policies, and the retrained DeepSWE-Verifier for CWM, whose workflow follows DeepSWE Agent.

\subsection{Learned scorer architecture and training}
\label{app:learned-scorer}
The learned scorer reads cached, mean-pooled hidden states rather than trajectory text. At every step it keeps the model-authored reasoning and function-call spans and excludes environment observations, following the assistant-only scoring scope of SWIFT~\citep{guo2026swift}. The spans are ordered as they occur in the trajectory. Let $u_{i,j}^{(\ell)}\in\mathbb{R}^{d}$ denote the mean-pooled state of span $j$ at layer $\ell$. We concatenate the states of all $L$ transformer layers without normalization:
\begin{equation}
 \tilde u_{i,j}=[u_{i,j}^{(1)};u_{i,j}^{(2)};\ldots;u_{i,j}^{(L)}]
 \in\mathbb{R}^{Ld}.
\end{equation}
A shared linear head maps each span to a gate logit and a local reward,
\begin{equation}
 (\tilde g_{i,j},v_{i,j})=W\tilde u_{i,j}+b,
 \qquad W\in\mathbb{R}^{2\times Ld},\quad b\in\mathbb{R}^{2},
 \label{eq:linear-head}
\end{equation}
and the trajectory logit is their gated average:
\begin{equation}
 q_{i,\mathrm{lin}}=
 \frac{\sum_j \sigma(\tilde g_{i,j})v_{i,j}}
 {\max\!\left(\sum_j \sigma(\tilde g_{i,j}),10^{-8}\right)}.
 \label{eq:linear-scorer}
\end{equation}
The head has $2Ld+2$ trainable parameters. We apply $W$ as one bias-free $d$-to-2 projection per layer, sum the outputs, and add the shared bias, which computes Equation~\ref{eq:linear-head} without forming the concatenated vector.

The training target of bank trajectory $i$ is its outcome label $y_i=\mathbf{1}[R_i\geq0.5]$, where $R_i$ is its rollout reward, and we minimize binary cross-entropy with $q_{i,\mathrm{lin}}$ as the logit. The bank is split into 80\% training and 20\% validation trajectories with random seed 0. We optimize with AdamW using a learning rate of $3\times10^{-6}$, weight decay of $10^{-5}$, and gradient accumulation over 16 trajectories for at most 40 epochs, and keep the checkpoint with the lowest validation loss. Each candidate trajectory is scored once with this checkpoint. Its logit is then converted to a normalized rank within the candidate set of its task instance (Appendix~\ref{app:rank-normalization}), which gives the learned score $s_{\mathrm{lin}}(\tau_i)$ of Equation~\ref{eq:scorer-fusion}.

\subsection{Token accounting details}
\label{app:r2e-token-accounting}
EF-verifier tokens sum the input and output tokens of the verifier calls actually made rather than estimating them from a global mean length. An average EF call uses 22.9K tokens over the full R2EGym-32B candidate pool and 18.9K tokens over the candidates that reach the final verifier. The corresponding DeepSWE-Preview averages are 51.7K and 46.2K tokens. The original cascade scores all 16 candidates, whereas \methodname calls the verifier only for the candidates that survive the execution stages, about four per task instance ($76.0\mathrm{K}/18.9\mathrm{K}\approx4.0$ on R2EGym-32B and $181.0\mathrm{K}/46.2\mathrm{K}\approx3.9$ on DeepSWE-Preview, Table~\ref{tab:cross-agent-main}). Most of the saving therefore comes from making fewer EF calls, and a smaller part comes from the shorter average length of the surviving trajectories.

We generate Stage~3 tests once for each of the 500 task instances with R2E-TestgenAgent~\citep{r2egym2025testgen}, the open-source test generator of the DeepSWE-Preview recipe, and run the same tests on the candidate patches of every configuration. A workflow is charged the test-generation cost only when it reaches Stage~3, at an average of 244.3K tokens per invocation according to the generation logs. On R2EGym-32B, Stage~3 is reached for 95.5\% of task instances in the original cascade and for 93.7\% in ours. The totals are therefore $366.2\mathrm{K}+0.955\times244.3\mathrm{K}\approx600\mathrm{K}$ and $76.0\mathrm{K}+0.937\times244.3\mathrm{K}\approx305\mathrm{K}$ tokens per task instance, the 49.1\% saving in Table~\ref{tab:cross-agent-main} and Figure~\ref{fig:cross-policy-saving}. The other configurations follow the same calculation with their own Stage~3 rates, for example 575.3K and 291.2K tokens on R2EGym-14B (Table~\ref{tab:cross-policy-best16}).

\subsection{Additional results}
\label{app:additional-results}
\paragraph{Smaller candidate budgets.}
Table~\ref{tab:cross-agent-bestk} extends the cross-agent comparison of Table~\ref{tab:cross-agent-main} to $K=4$, 8, and 12 under the same protocol. The original cascade of each agent uses the policy-associated verifier in both EF stages. At every budget, \methodname stays within 0.81 points of this cascade, and at $K=16$ it matches or exceeds it on all three agents, while its token saving widens with $K$ (Figure~\ref{fig:main-results}).

\begin{table}[!htbp]
\caption{Hybrid Best@$K$ (\%) of the cross-agent comparison at $K=4$, 8, 12, and 16. The $K=16$ column repeats Table~\ref{tab:cross-agent-main}.}
\label{tab:cross-agent-bestk}
\centering
\footnotesize
\setlength{\tabcolsep}{3pt}
\resizebox{\linewidth}{!}{%
\begin{tabular}{lcccccccccccc}
\hline\noalign{\vskip 1.5pt}
& \multicolumn{4}{c}{DeepSWE Agent} & \multicolumn{4}{c}{R2EGym Agent} & \multicolumn{4}{c}{CWM Agent}\\
Stage~1 verifier & $K{=}4$ & $K{=}8$ & $K{=}12$ & $K{=}16$ & $K{=}4$ & $K{=}8$ & $K{=}12$ & $K{=}16$ & $K{=}4$ & $K{=}8$ & $K{=}12$ & $K{=}16$\\
\hline\noalign{\vskip 1.5pt}
DeepSWE-Verifier & 53.99 & 57.10 & 58.43 & 59.26 & 42.65 & 46.11 & 48.23 & 49.70 & 52.13 & 52.93 & 53.80 & 54.21\\
R2EGym-Verifier & 53.67 & 56.38 & 57.59 & 57.95 & 42.35 & 45.04 & 46.27 & 46.56 & 52.20 & 53.28 & 53.77 & 53.51\\
DEV matching-pairs & 53.71 & 56.70 & 57.82 & 58.75 & 42.49 & 45.58 & 47.38 & 48.38 & 52.13 & 52.66 & 53.07 & 53.11\\
AgentPRM & 53.55 & 56.70 & 58.40 & 59.76 & 41.47 & 44.22 & 45.74 & 46.56 & 51.96 & 52.58 & 52.85 & 53.31\\
\textbf{\methodname (ours)} & 53.45 & 56.92 & 58.64 & 60.06 & 41.79 & 44.23 & 46.12 & 47.73 & 51.75 & 52.65 & 53.60 & 54.21\\
\noalign{\vskip 1pt}\hline
\end{tabular}}
\end{table}

\paragraph{Cross-policy comparison.}
Table~\ref{tab:cross-policy-best16} lists the resolution rates and token costs behind Figure~\ref{fig:cross-policy-saving}. \methodname reduces EF-verifier tokens by 79.2\% on R2EGym-32B and by 81.0\% on R2EGym-14B, and total verification tokens by 49.1\% and 49.4\%, respectively. Hybrid Best@16 rises by 1.17 points on R2EGym-32B and stays within 0.33 points on R2EGym-14B.

\begin{table}[!htbp]
\caption{Cross-policy results on R2EGym Agent at $K=16$. Best@16 is the resolution rate (\%). Token cost gives EF-verifier (EF) and total verification tokens per task instance, in thousands. Both cascades use R2EGym-Verifier in Stage~4.}
\label{tab:cross-policy-best16}
\centering
\footnotesize
\setlength{\tabcolsep}{5pt}
\begin{tabular}{lrrrrrr}
\hline\noalign{\vskip 1.5pt}
& \multicolumn{3}{c}{R2EGym-32B} & \multicolumn{3}{c}{R2EGym-14B}\\
\multirow{2}{*}{Stage~1 verifier} & \multirow{2}{*}{Best@16} & \multicolumn{2}{c}{Token cost (K)} & \multirow{2}{*}{Best@16} & \multicolumn{2}{c}{Token cost (K)}\\
\cline{3-4}\cline{6-7}\noalign{\vskip 0.8pt}
& & EF & Total & & EF & Total\\
\hline\noalign{\vskip 1.5pt}
R2EGym-Verifier & 46.56 & 366.2 & 599.6 & 43.28 & 344.2 & 575.3\\
\textbf{\methodname (ours)} & \textbf{47.73} & \textbf{76.0} & \textbf{304.9} & \textbf{42.95} & \textbf{65.5} & \textbf{291.2}\\
\noalign{\vskip 1pt}\hline
\end{tabular}
\end{table}

\end{document}